\documentclass[reprint,superscriptaddress,nofootinbib,twocolumn,amsmath,amssymb,aps,prl]{revtex4-2}

\usepackage{graphicx}
\usepackage{dcolumn}
\usepackage{float}
\usepackage{bm}
\usepackage[colorlinks]{hyperref}
\usepackage[utf8]{inputenc}	
\usepackage{lmodern} 
\usepackage{siunitx}
\usepackage[version=4]{mhchem}
\usepackage{microtype}
\usepackage{diagbox}

\begin{document}
\title{Multiphoton Quantum Imaging using Natural Light}

\author{Fatemeh Mostafavi}
\affiliation{Quantum Photonics Laboratory, Department of Physics \& Astronomy, Louisiana State University, Baton Rouge, LA 70803, USA}

\author{Mingyuan Hong}
\affiliation{Quantum Photonics Laboratory, Department of Physics \& Astronomy, Louisiana State University, Baton Rouge, LA 70803, USA}

\author{Riley B. Dawkins}
\affiliation{Quantum Photonics Laboratory, Department of Physics \& Astronomy, Louisiana State University, Baton Rouge, LA 70803, USA}

\author{Jannatul Ferdous}
\affiliation{Quantum Photonics Laboratory, Department of Physics \& Astronomy, Louisiana State University, Baton Rouge, LA 70803, USA}

\author{Rui-Bo Jin}
\affiliation{Hubei Key Laboratory of Optical Information and Pattern Recognition, Wuhan Institute of Technology, Wuhan 430205, China}

\author{Roberto de J. Le\'on-Montiel}
\affiliation{Instituto de Ciencias Nucleares, Universidad Nacional Aut\'onoma de M\'exico, Apartado Postal 70-543, 04510 Cd. Mx., M\'exico}

\author{Chenglong You}
\email{cyou2@lsu.edu}
\affiliation{Quantum Photonics Laboratory, Department of Physics \& Astronomy, Louisiana State University, Baton Rouge, LA 70803, USA}

\author{Omar S. Maga\~na-Loaiza}
\affiliation{Quantum Photonics Laboratory, Department of Physics \& Astronomy, Louisiana State University, Baton Rouge, LA 70803, USA}

\date{\today}

\begin{abstract}
It is thought that schemes for quantum imaging are fragile against realistic environments in which the background noise is often stronger than the nonclassical signal of the imaging photons. Unfortunately, it is unfeasible to produce brighter quantum light sources to alleviate this problem. Here, we overcome this paradigmatic limitation by developing a quantum imaging scheme that relies on the use of natural sources of light. This is achieved by performing conditional detection on the photon number of the thermal light field scattered by a remote object. Specifically, the conditional measurements in our scheme enable us to extract quantum features of the detected thermal photons to produce quantum images with improved signal-to-noise ratios. This technique shows a remarkable exponential enhancement in the contrast of quantum images.  Surprisingly, this measurement scheme enables the possibility of producing images from the vacuum fluctuations of the light field. This is experimentally demonstrated through the implementation of a single-pixel camera with photon-number-resolving capabilities. As such, we believe that our scheme opens a new paradigm in the field of quantum imaging. It also unveils the potential of combining natural light sources with nonclassical detection schemes for the development of robust quantum technologies.

\end{abstract}

\maketitle

The use of nonclassical correlations of photons to produce optical images in a nonlocal fashion gave birth to the field of quantum imaging almost three decades ago \cite{PhysRevApticalimaging, PhysRevLettTwoPhotonCoincidence, Magana-Loaiza_2019}. Interestingly, it was then discovered that exploiting the quantum properties of the light field enables improving the resolution of optical instruments beyond the diffraction limit \cite{TwoPhotonLithography, PhysRevLettScalableSpatial, PhysRevLettDowling, PhysRevXSuperresolution, npj2022}. It was also shown that schemes for quantum imaging allow for the formation of images with sub-shot-noise levels of precision \cite{Brida2010NaturePhotonics, Quantumsecuredimaging, Lemosundetectedphotons}. These features have been exploited to demonstrate the formation of few-photon images with high contrast \cite{APlMagana-Loaiza2013, Morris2015NatureComm, BarretoLemosmetrology}.  Furthermore, the compatibility of quantum imaging techniques with protocols for quantum cryptography have cast interest in the development of schemes for quantum-secured imaging \cite{Quantumsecuredimaging, Magana-Loaiza_2019}. Despite the enormous potential of quantum imaging for microscopy, remote sensing, and astronomy, schemes for quantum imaging remain fragile against realistic conditions of loss and noise \cite{ref1Microscopy, PhysRevAGhostimaging, PhysRevLettPlenopticImaging, Magana-Loaiza_2019, Genovese}. Unfortunately, these limitations render the realistic application of quantum imaging unfeasible \cite{ref1Sciencespatialcorrelations, Magana-Loaiza_2019}.

Sharing similarities with other quantum technologies, existing techniques for quantum imaging rely on the use of nonclassical states of light \cite{OBrien2009NaturePhotonics, Lawrie2019SqueezedLight, midgal2013}. However, the brightness of available quantum light sources is generally low \cite{Multiphotonengineeri, midgal2013, Hong2023}. For example, existing sources of nonclassical light allow for the preparation of few-photon states that exhibit fragile quantum correlations \cite{PhysRevLettThermalLight,PhysRevLett.TurbulenceFree, ref1Sciencespatialcorrelations}. This situation leads to common scenarios where environmental noise is typically larger than the signal of photons produced by processes of spontaneous parametric down-conversion or four-wave mixing \cite{DELLANNO200653, midgal2013}. Unfortunately, it is not feasible to produce brighter quantum light sources to overcome these limitations. Moreover, losses and noise cannot be avoided in realistic scenarios \cite{Magana-Loaiza_2019}. Thus, any robust protocol for quantum imaging must rely on ubiquitous natural sources of light, such as thermal light. 

Here we demonstrate the extraction of quantum images from classical noisy images produced by thermal sources of light \cite{You2023CittertZernike, PhysRevAconditionalmeasurements}.  Specifically, our quantum imaging scheme isolates multiphoton subsystems of thermal light sources to dramatically improve the signal-to-noise ratio of imaging instruments. This robust protocol for quantum imaging is demonstrated through the implementation of a novel single-pixel camera with photon-number-resolving capabilities \cite{You2020Gerritsfication}. Surprisingly, this quantum camera enables the extraction of information from the vacuum-fluctuation components of thermal light sources to produce quantum images with improved contrast. This technique shows a remarkable exponential improvement in the contrast of quantum images.
We also demonstrate the possibility of using correlated multiphoton subsystems to form high-contrast quantum images from images in which the background noise is comparable to the signal of thermal light sources. These surprising results can only be explained using quantum physics \cite{PhysRevCoherentIncoherent, PhysRevLettSudarshan}. Our work unveils the potential of combining natural light with nonclassical detection schemes for the development of robust quantum technologies. We believe that our findings open a new paradigm in the field of quantum imaging. 

\begin{figure*}[!tbp]
\centering	
\includegraphics[width=0.85\textwidth]{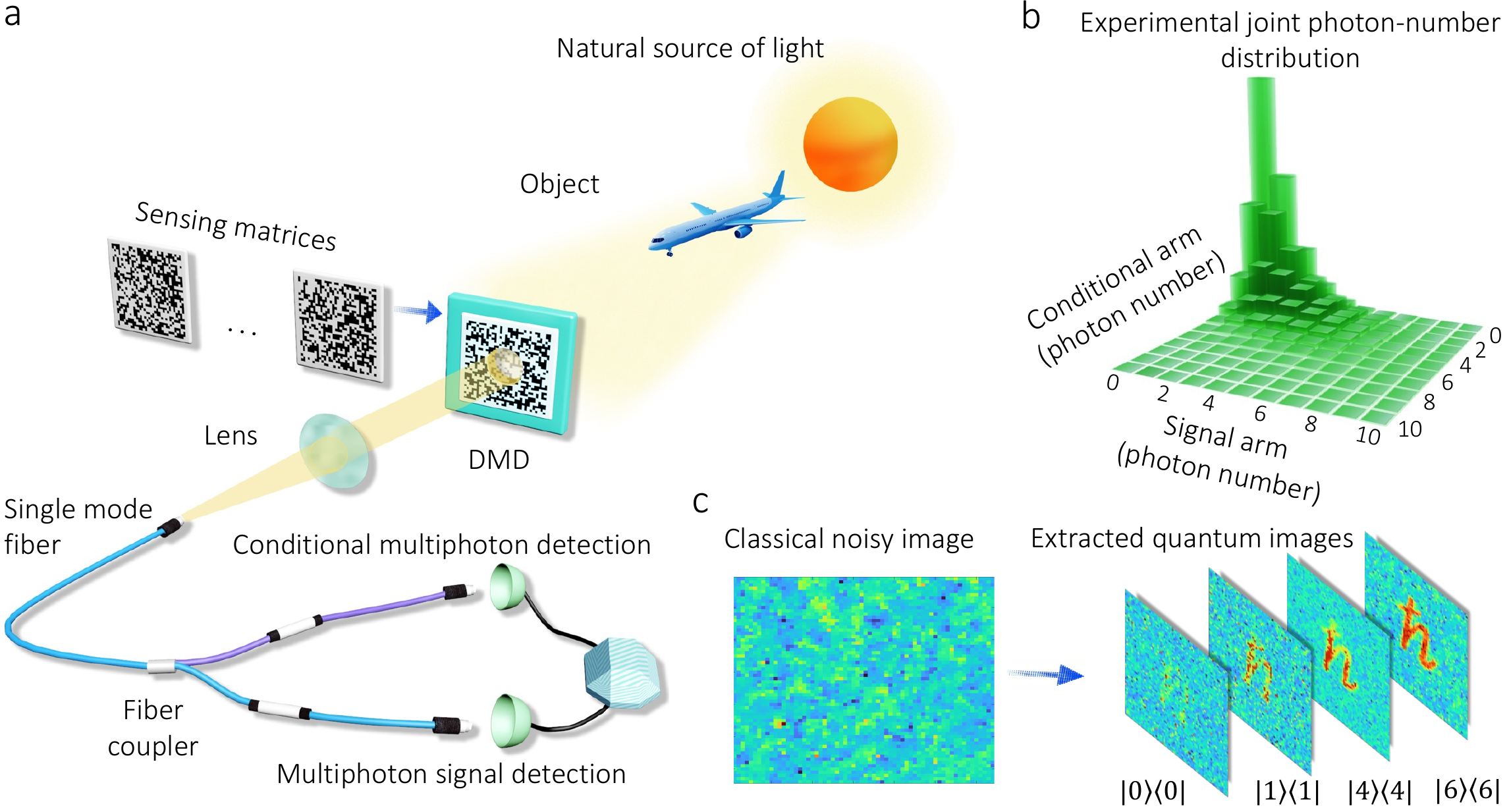}
	\caption{\textbf{Multiphoton quantum imaging using natural sources of light.} 
The schematic in \textbf{a} depicts the implementation of a quantum camera with photon-number-resolving (PNR) capabilities. Here the thermal light field reflected off a target object is projected into a series of random binary matrices and then coupled into a single-mode fiber (SMF). The binary sensing matrices are displaced onto a digital micromirror device (DMD). Further, the thermal light field coupled into the SMF  is split by a 40:60 fiber coupler and measured by two PNR detectors. We report the experimental joint photon-number distribution of our thermal light source in \textbf{b}. In this case, the degree of second-order coherence, $g^{(2)}(0)$, of the thermal light source is equal to 2. The series of PNR measurements for different binary sensing matrices enables us to use compressive sensing (CS) to demonstrate a single-pixel camera with PNR capabilities. As shown in \textbf{c}, our ability to measure the multiphoton subsystems, represented by the elements of the joint photon-number distribution of the thermal source, enables us to demonstrate quantum imaging even in situations in which noise prevents the formation of the classical image of the object. Specifically, the environmental noise in \textbf{c} forbids the imaging of the character $\hbar$. However, the projection of the thermal light field into its vacuum  component reveals the presence of the object. Remarkably, the projection into larger multiphoton subsystems enables the extraction of quantum images of the object that was not visible in the classical image.} 
	\label{fig:figure1}
 \end{figure*}

In Fig. \ref{fig:figure1}\textbf{a} we illustrate the experimental implementation of our scheme for multiphoton quantum imaging. Here, the thermal light reflected off a target object is projected onto a digital micromirror device (DMD) where a series of random binary patterns are displayed. The thermal photons from the DMD are collected by a single-mode fiber (SMF) and then probabilistically split by a fiber coupler. The photons in each fiber are measured by two photon-number-resolving (PNR) detectors \cite{You2020Gerritsfication, npj2022}. The random sensing matrices displayed on the DMD are used to implement a single-pixel camera \cite{PRLMirhossein2014, tutorial, Montaut2018, PhysRevAGhostimaging}. Further, our photon counting scheme enables us to project the coupled thermal light field into its constituent multiphoton subsystems. The joint photon-number distribution of the thermal source is reported in Fig. \ref{fig:figure1}\textbf{b}. The classical nature of the source is certified by the degree of second-order coherence $g^{(2)}$, which is equal to 2 in our experiment \cite{Gerry2004}. Each element in this joint probability distribution represents a multiphoton subsystem that we can isolate through the implementation of projective measurements \cite{PRLMirhossein2014, tutorial, Montaut2018, PhysRevAGhostimaging}. This measurement approach lies at the hearth of our protocol for multiphoton quantum imaging. 

\begin{figure*}[!tbp]
\centering	
\includegraphics[width=0.87\textwidth]{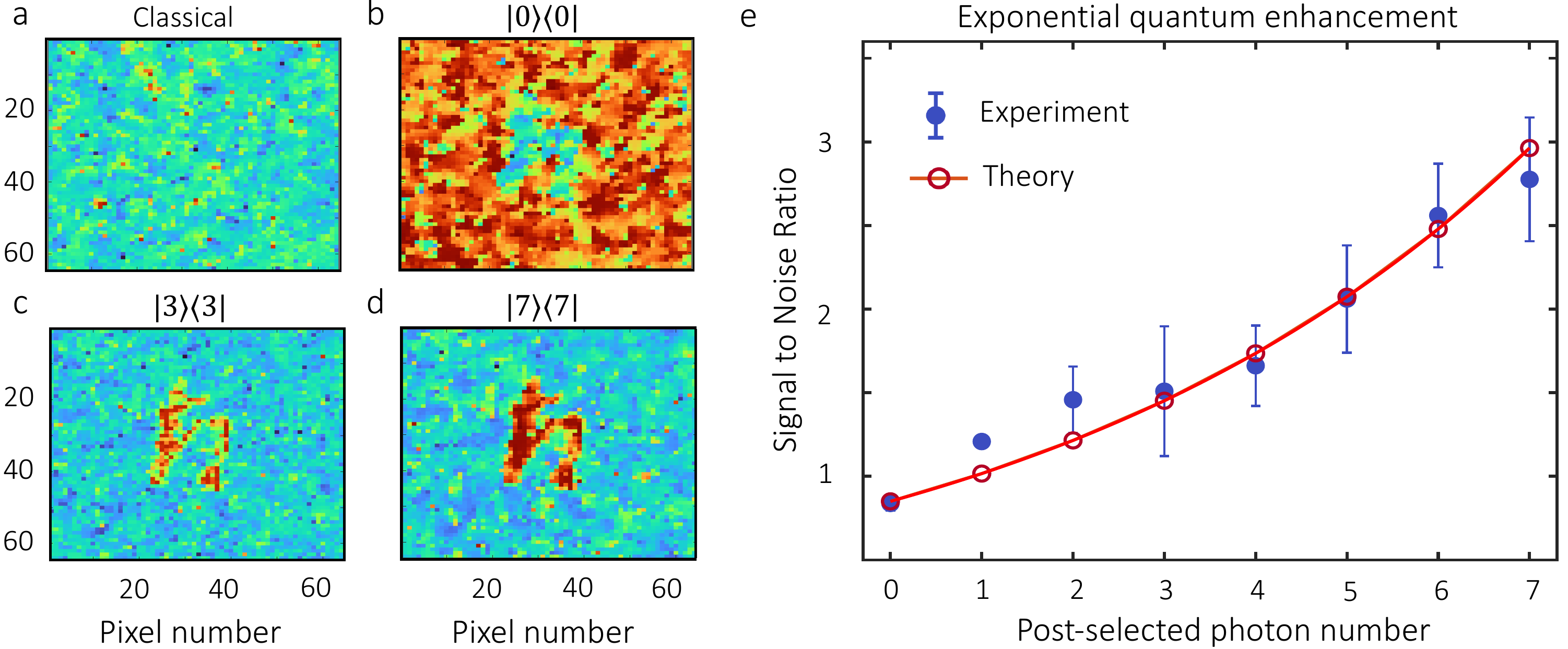}
\caption{\textbf{Extraction of quantum images from a classical CS reconstruction.} The reconstructed image using our single-pixel camera for classical thermal light is shown in \textbf{a}. In this case, environmental noise is higher than the signal and consequently the reconstructed image shows a low contrast that prevents the observation the object. Surprisingly, the projection of the light field into its vacuum component boosts the contrast of the image, this is reported in \textbf{b}. Naturally, the formation of this image cannot be understood using classical optics. As demonstrated in \textbf{c}, the projection of the thermal source of light into three-photon events enables the extraction of a quantum image with an improved signal-to-noise ratio (SNR). Remarkably, the projection of the detected thermal field into seven-particle subsystems leads to the formation of the high-contrast quantum image in \textbf{d}. As reported in \textbf{e}, and in agreement with Eq. (\ref{Eq:post}), the improvement in the SNR is exponential with the number of projected photons.  These results were obtained using 25$\%$ of the total number of measurements that can be used in our CS algorithm. Furthermore, the mean photon number $\bar{n}_t$ of the thermal light source is 0.8.} 
\label{fig:figure2}
\end{figure*}

As shown in Fig. \ref{fig:figure1}\textbf{c}, the projection of thermal light scattered by a target object into its constituent multiphoton subsystems enables the formation of high-contrast quantum images. This surprising effect enables extracting quantum images of a target object, even when environmental noise prevents the formation of its classical image through intensity measurements. We now describe the quantum multiphoton processes that make this effect possible. For the sake of simplicity, we assume the uniform illumination of the object $\Vec{\boldsymbol{s}}_0$ by a thermal light field. As depicted in Fig. \ref{fig:figure1}\textbf{a}, 
the projection of the object into random sensing matrices, represented by the covector $\Vec{\boldsymbol{Q}}_t$, enables us to discretize the object into $X$ pixels. The label $t$ indexes the different sensing matrices. All such matrices can be represented by the $M\times X$ matrix  $\boldsymbol{Q} = \bigoplus_{t=1}^M \Vec{\boldsymbol{Q}}_t$, where $M$ is the number of sensing matrix configurations. Then, each filtering configuration results in a thermal state with a mean photon number given by $\bar{n}_t = \Vec{\boldsymbol{Q}}_t\cdot \Vec{\boldsymbol{s}}_0$. The multiphoton state after the fiber coupler can be written in terms of the Glauber-Sudarshan $P$ function as \cite{PhysRevLettSudarshan,PhysRevCoherentIncoherent}
\begin{equation}
\begin{aligned}
\label{Eq:state}
      \hat{\boldsymbol{\rho}}_{\boldsymbol{Q}} =&
      \bigoplus_{t=1}^M\int d^2\alpha \frac{1}{\pi\bar{n}_t}e^{-\frac{\left|\alpha\right|^2}{\bar{n}_t}}\\
      &\times\left|\alpha \cos(\theta),i\alpha\sin(\theta)\rangle\langle\alpha \cos(\theta),i\alpha\sin(\theta)\right|_{a,b}.
\end{aligned}
\end{equation}
The indices $a$ and $b$ denote the output modes of the fiber coupler. Furthermore, the parameter $\theta$ describes the splitting ratio between the two output ports. 

Next, we describe the signal-to-noise ratio (SNR) and how this quantity is modified by projecting the thermal field into its constituent multiphoton subsystems. To account for noise, we must consider photocounting with quantum efficiencies $\eta_{a/b}$ and noise counts $\nu_{a/b}$ \cite{Multiphotonengineeri,PhysRevATruephoton,PhysRevAmultimodethermal}. Specifically, the joint photon-number distribution reported in Fig. \ref{fig:figure1}\textbf{b}, can be mathematically described as
\begin{widetext}
   \begin{equation}
   \begin{aligned}
   \label{Eq:joint}
               \Vec{\boldsymbol{p}}_{\boldsymbol{Q}}(n, m)&=\bigoplus_{t=1}^M\left\langle: \frac{\left(\eta_a \hat{n}_a+\nu_a\right)^n}{n !} e^{-\left(\eta_a \hat{n}_a+\nu_a\right)} \otimes \frac{\left(\eta_b \hat{n}_b+\nu_b\right)^m}{m !} e^{-\left(\eta_b \hat{n}_b+\nu_b\right)}:\right\rangle\\
               &=\bigoplus_{t=1}^M\frac{e^{-\nu_a-\nu_b}}{\bar{n}_t n! m!}\sum_{i=0}^n\sum_{j=0}^m \binom{n}{i}\binom{m}{j}(i+j)!\frac{\eta_a^i\eta_b^j\nu_a^{n-i}\nu_b^{m-j}}{\left(\frac{1}{\bar{n}_t}+\eta_a\cos^2(\theta)+\eta_b\sin^2(\theta)\right)^{1+i+j}}\cos^{2i}(\theta)\sin^{2j}(\theta),
   \end{aligned}
    \end{equation}  
\end{widetext}
\noindent
where $\hat{n}_{a/b}$ is the photon number operator, and $:\cdot:$ represents the normal ordering prescription. We write the $t^{\text{th}}$ component of this vector as $p_{\boldsymbol{Q},t}(n,m)$. Additionally, when there is no signal and only noise is measured, we will have the probability distribution $p_{n,i}(k) = e^{-\nu_i}\frac{\nu_i^k}{k!}$ in each arm, where $i = a,b$. The joint probability distribution in this case is then given by $p_n(k,l) = p_{n,a}(k)p_{n,b}(l)$.
 
The two-mode multiphoton system described by Eq. (\ref{Eq:joint}) enables two schemes for projective measurements that lead to different scaling factors for the SNR of quantum images. First, we project one of the arms into a particular multiphoton subsystem. In other words, we ignore arm $b$ and implement a photon-number-projective measurement in arm $a$. For such post-selection on a multiphoton subsystem with $N$ photons, the SNR scales with 

\begin{equation}
      \label{Eq:post}
      \overrightarrow{\textbf{SNR}}_{\text{post}} = \frac{\sum_{m=0}^\infty \Vec{\boldsymbol{p}}_{\boldsymbol{Q}}(N,m)}{p_{n,a}(N)}=\frac{\Vec{\boldsymbol{p}}_{\boldsymbol{Q}}(N)}{p_{n,a}(N)}.
\end{equation} 
Remarkably, this expression follows an exponentially increasing trend with respect to $N$, meaning that post-selection can significantly reduce the noise of a quantum image.

\begin{figure*}[!tbp]
\centering	
\includegraphics[width=0.9\textwidth]{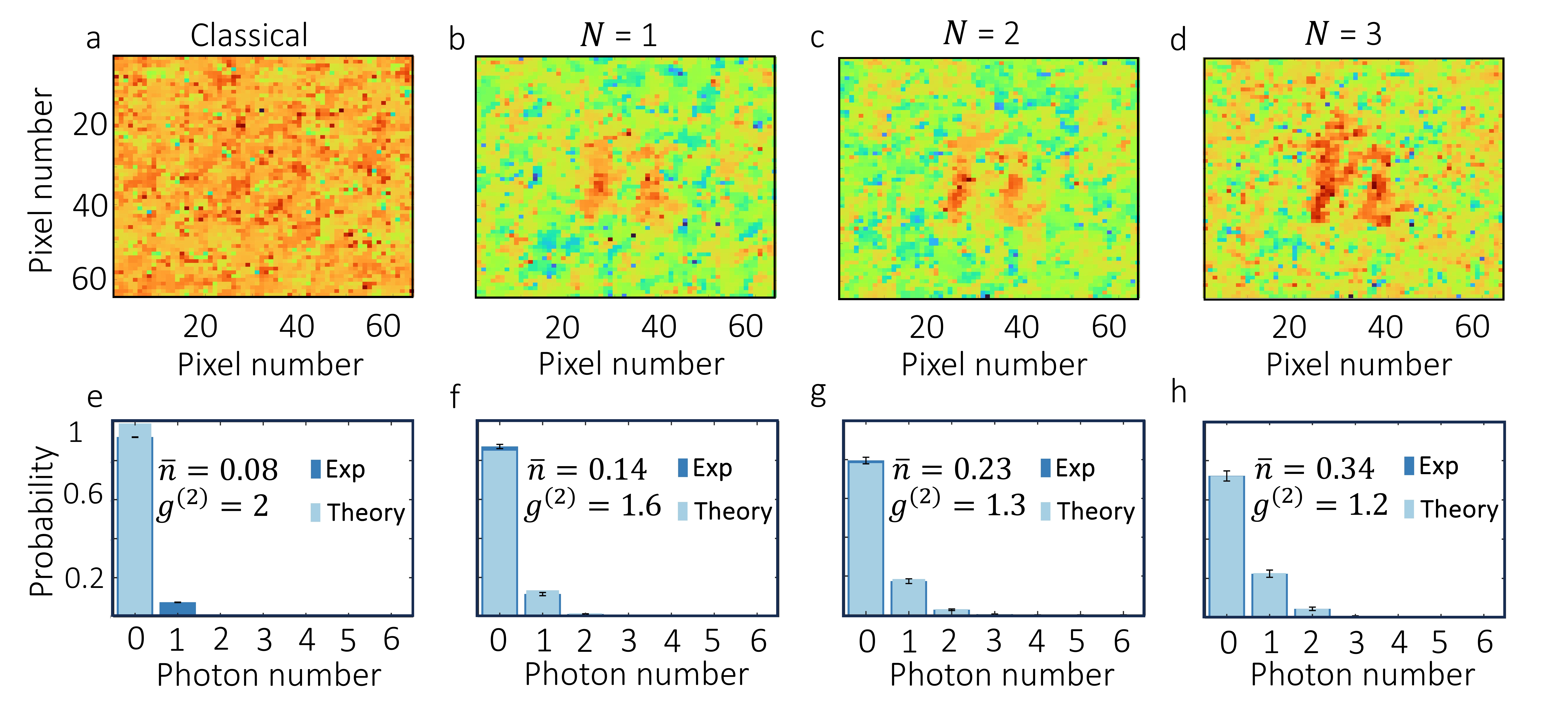}
\caption{\textbf{Photon-subtracted multiphoton quantum imaging.} The noise accompanying a signal reflected off a target object produces the classical image reported in \textbf{a}. Here, it is not possible to identify the object of interest with a classical single-pixel camera \cite{APlMagana-Loaiza2013}. The mean photon number $\bar{n}_t$ of our thermal light source is $0.08$. Interestingly, the subtraction of one photon improves the contrast of the image leading to the CS reconstruction in \textbf{b}. Furthermore, our single-pixel camera with PNR capabilities enables multiphoton subtraction to produce the quantum images shown in  \textbf{c} and \textbf{d}. In these cases, we subtracted two and three photons, respectively. These images were produced using only 12$\%$ of the total number of measurements that can be used in our CS algorithm. The advantage provided by our protocol for photon-subtracted quantum imaging can be understood through the photon-number distributions reported from \textbf{e} to \textbf{h}. The unconditional detection of the weak thermal light signal produces the histogram in \textbf{e}. This histogram unveils the overwhelming presence of vacuum and single-photon events used to reconstruct the image in \textbf{a}. Furthermore, as shown in \textbf{f}, the process of one-photon subtraction increases the mean photon number of the thermal signal while reducing its degree of second-order coherence $g^{(2)}$. The subtraction of two-photon events leads to a stronger signal characterized by the histogram in \textbf{g}. This conditional signal produces the enhanced image of the object in \textbf{c}. Notably, the implementation of three-photon subtraction leads to the optical signal with nearly coherent statistics reported in \textbf{h}. This boosted signal enables the reconstruction of the high-contrast image in \textbf{d}. } 
\label{fig:figure3}
\end{figure*}

The second approach relies on the subtraction of $N$ photons from the thermal multiphoton system in Eq. (\ref{Eq:joint}) \cite{PhysRevAconditionalmeasurements, Mostafavinanophotonic, HashemiRafsanjani2017}. This procedure entails measuring photon events in arm $a$ conditioned on the detection of $N$ photons in arm $b$. Using Eq. (\ref{Eq:joint}), the intensity in arm $a$ is then given by $\langle\hat{\boldsymbol{n}}_a\rangle_N = \bigoplus_{t=0}^M\left(\sum_{k=0}^\infty k p_{\boldsymbol{Q},t}(k,N)\right)/\left(\sum_{k=0}^\infty p_{\boldsymbol{Q},t}(k,N)\right)$. Additionally, the photon-subtracted noise can be written as $\langle\hat{n}_a\rangle_{N,0} = \bigoplus_{t=0}^M\left(\sum_{k=0}^\infty k p_n(k,N)\right)/\left(\sum_{k=0}^\infty p_n(k,N)\right)$. This scheme leads to the following expression for the SNR:
\begin{equation}
 \label{Eq:sub}
    \overrightarrow{\textbf{SNR}}_{\text{sub}} = \frac{\langle\hat{\boldsymbol{n}}_a\rangle_N}{\langle\hat{n}_a\rangle_{N,0}}.
\end{equation}
The quantum enhancement for the SNR in this case is linearly increasing with respect to $N$. Therefore, photon-subtraction is also an effective means for noise-reduction.

The series of spatial projective measurements described by the vector $\Vec{\boldsymbol{Q}}_t$ enables implementing a single-pixel camera with photon-number resolving capabilities via compressive sensing (CS) \cite{PRLMirhossein2014, tutorial, Montaut2018, PhysRevAGhostimaging}. This technique permits the reconstruction of  multiphoton quantum images described by $\Vec{\boldsymbol{s}}'$ via the minimization of the following quantity with respect to $\Vec{\boldsymbol{s}}'$: 
\begin{equation}
\label{eq:cs}
     \sum_{i=0}^X \lVert\nabla s_i'\rVert_{l_1} + \frac{\mu}{2}\lVert \boldsymbol{Q}\Vec{\boldsymbol{s}}' - \langle\hat{\boldsymbol{n}}\rangle\rVert_{l_2}.
\end{equation}
As described above, $\langle\hat{\boldsymbol{n}}\rangle$ could be either $\Vec{\boldsymbol{p}}_{\boldsymbol{Q}}(N)$ or $\langle\hat{\boldsymbol{n}}_a\rangle_N$. Moreover, the $1$- and $2$-norm are denoted by $\lVert\cdot\rVert_{l_1}$ and $\lVert\cdot\rVert_{l_2}$, respectively. The discrete gradient operator is described by $\nabla$, and the penalty factor by $\mu$  \cite{tutorial, PRLMirhossein2014,Montaut2018, PhysRevAGhostimaging}. 


We now discuss the experimental process of quantum-image extraction from classical images. This was implemented using one PNR detector. In Fig. \ref{fig:figure2}\textbf{a}, we show the CS reconstruction of a classical image for a situation in which environmental noise is comparable to the signal. In this case, the level of noise forbids the observation of the object. The mean photon number $\bar{n}_{t}$ of the thermal light source is 0.8.  Surprisingly, the projection of the thermal signal into its vacuum component reveals the presence of the object. As such, the quantum image in Fig. \ref{fig:figure2}\textbf{b} is formed by the vacuum-fluctuation component of the electromagnetic field and cannot be explained using classical physics \cite{Genovese,PhysRevLettSudarshan,PhysRevCoherentIncoherent, You2023CittertZernike}. This nonclasical reconstruction, obtained from the measurement of vacuum events, demonstrates that the process of projecting the thermal light signal into one of its constituent quantum subsystems, such as the vacuum, modifies the SNR as established by Eq. (\ref{Eq:post}). As suggested by the reconstruction in Fig. \ref{fig:figure2}\textbf{c}, the post-selection on higher multiphoton events leads to quantum images with an improved contrast. Interestingly, the projection of the thermal light signal into seven-photon subsystems leads to a dramatic improvement of the contrast of the image. This effect becomes evident in the quantum image shown in Fig. \ref{fig:figure2}\textbf{d}. Remarkably, the exponential growth of the SNR with the number of projected multiphoton subsystems is summarized in Fig. \ref{fig:figure2}\textbf{e}. These results demonstrate that our single-pixel camera with PNR capabilities enables the extraction of quantum multiphoton images with unprecedented degrees of contrast \cite{Brida2010NaturePhotonics,npj2022, Morris2015NatureComm, Genovese, Magana-Loaiza_2019}. 

\begin{figure}[!tbp]
\includegraphics[width=\linewidth]{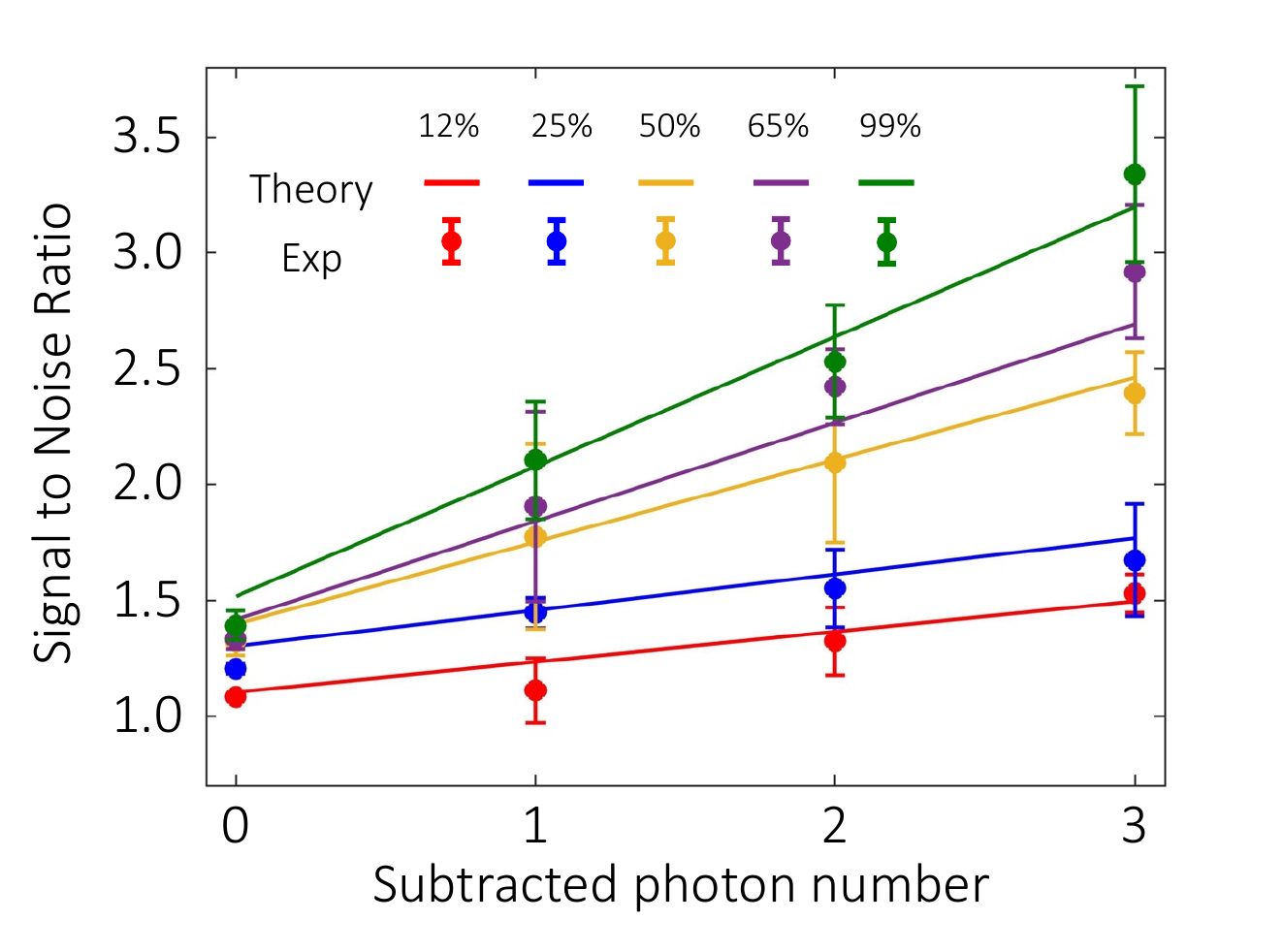}
\caption{\textbf{Performance of photon-subtracted multiphoton quantum imaging.} The SNR of the photon-subtracted quantum images shows a linear dependence on the number of subtracted photons. This behavior is in good agreement with Eq. (\ref{Eq:sub}). Interestingly, the collection of larger sets of data leads to faster improvements of the SNR for our multiphoton quantum imaging scheme. } 
\label{fig:figure4}
\end{figure}

While the projection of thermal light into its constituent multiphoton subsystems enables the extraction of quantum images with high contrast, it is also possible to correlate photon events to improve the SNR of a quantum imaging protocol. This feature also enables us to perform quantum imaging at low light levels. We now experimentally demonstrate this possibility by implementing a scheme for photon subtraction on our single-pixel camera with PNR capabilities. In this case, the mean photon number $\bar{n}_t$ is equal to 0.08,
one order of magnitude lower than the brightness of the source used for the experiment discussed in Fig. \ref{fig:figure2}. As illustrated in Fig. \ref{fig:figure1}\textbf{a}, this quantum imaging scheme utilizes two PNR detectors \cite{Multiphotonengineeri,You2021Scalablemultiphoton}. First, we use the noisy thermal signal to reconstruct the classical image shown in Fig. \ref{fig:figure3}\textbf{a}. Here, the large levels of noise forbid the observation of the target object. Remarkably, the subtraction of one photon from the thermal noisy signal reveals the presence of the object in Fig. \ref{fig:figure3}\textbf{b}. As predicted by Eq. (\ref{Eq:sub}), the process of multiphoton subtraction leads to enhanced quantum images. Specifically, two-photon subtraction leads to the improved image in Fig. \ref{fig:figure3}\textbf{c}. Furthermore, the CS reconstruction of the three-photon subtracted quantum image reported in Fig. \ref{fig:figure3}\textbf{d} shows a significant  improvement of the contrast with respect to the classical image in Fig. \ref{fig:figure3}\textbf{a}. The physics behind our scheme for quantum imaging can be understood through the increasing mean photon number that characterizes the histograms shown from Fig. \ref{fig:figure3}\textbf{e} to \textbf{h}. Moreover, the thermal fluctuations of the detected field are reduced by subtracting photons \cite{PhysRevAconditionalmeasurements, Magana-Loaiza_2019, HashemiRafsanjani2017}. This effect is indicated by the decreasing degree of second-order coherence $g^{(2)}$ corresponding to the photon-number distributions in Fig. \ref{fig:figure3}.

The improvement in the SNR of the experimental photon-subtracted quantum images is quantified in Fig. \ref{fig:figure4}. In agreement with Eq. (\ref{Eq:sub}), the contrast of the filtered images, as a function of the number of subtracted photons, follows a linear dependence. Although, the benefits of our photon-subtracted scheme for multiphoton quantum imaging are evident for small and incomplete sets of data, the rate at which the SNR increases can be further amplified by collecting larger sets of data. It is worth noting that the exponential and linear mechanisms, reported in Fig. \ref{fig:figure2} and Fig. \ref{fig:figure4}, for improving the SNR of weak and noisy imaging signals have the potential to enable the realistic application of robust quantum cameras with PNR capabilities \cite{npj2022,Cheng2023}. As such, these findings could lead to novel quantum techniques for multiphoton microscopy and remote sensing \cite{Morris2015NatureComm, Magana-Loaiza_2019, Genovese}.

Quantum imaging schemes have been demonstrated to be fragile against realistic environments in which the background is comparable to the nonclassical signal of the imaging photons 
\cite{ref1Microscopy, PhysRevAGhostimaging, PhysRevLettPlenopticImaging, Magana-Loaiza_2019, Genovese, ref1Sciencespatialcorrelations, Magana-Loaiza_2019}. This issue prevents the realistic application of quantum imaging techniques for microscopy, remote sensing, and astronomy \cite{Morris2015NatureComm, Magana-Loaiza_2019, Genovese}. In this work, we overcome this paradigmatic limitation by developing a multiphoton quantum imaging scheme that relies on the use of natural sources of light.  This is demonstrated through the implementation of a single-pixel camera with photon number resolving capabilities that enables the projection of classical thermal light fields into its constituent multiphoton subsystems. This kind of quantum measurement enables us to extract high-contrast quantum images from noisy classical images of target objects. Our technique shows a remarkable exponential enhancement in the contrast of quantum images.
Surprisingly, we demonstrated the formation of quantum images produced by the vacuum-fluctuation components of thermal light sources. We also demonstrate the possibility of using correlated multiphoton subsystems to form high-contrast quantum images from images in which the background noise is comparable to the signal of thermal light sources. Thus, we believe that our scheme opens a new paradigm in the field of quantum imaging \cite{Magana-Loaiza_2019, Genovese,OBrien2009NaturePhotonics, Lawrie2019SqueezedLight}. Furthermore, it unveils the potential of combining natural light sources with nonclassical detection schemes for the development of robust quantum technologies \cite{Magana-Loaiza_2019, Genovese,OBrien2009NaturePhotonics, Lawrie2019SqueezedLight, DELLANNO200653}.

\section{Methods}
\subsection{Experimental Setup}
Our proof-of-principle quantum imaging setup utilizes pseudo-thermal light, which has the same properties of coherence as natural sources of light \cite{Arecchi65PRL}. This source is generated by passing the coherent light from a continuous-wave laser at 633 nm through a rotating ground glass \cite{Arecchi65PRL,You2020Gerritsfication}. The thermal light is then collected into a single-mode fiber and collimated with a lens ($f=5$ cm) to illuminate the target object. Here, the target object ``$\hbar$'' is generated using a digital micro-mirror device (DLP6500 DLP$\textsuperscript{\textregistered}$ DMD). Then, the reflected ``$\hbar$'' is projected onto a second DMD with a 4-f system comprised of two lenses, each with a focal length of 10 cm. This second DMD facilitates compressive sensing by displaying a series of random binary matrices \cite{PRLMirhossein2014, APlMagana-Loaiza2013}. Next, the reflected light from this DMD is imaged using a another 4-f system comprised of two lenses, with focal lengths of 25 and 10 cm. Then, we couple the reflected light into a 1$\times$2 50:50 fiber beam splitter (Thorlabs TW630R5F1) using a Rochester lens ($f=4.5$ mm). The split beams are detected by two fiber-coupled avalanche photodiodes (APDs, Excelitas SPCM-AQRH-13-FC), where photon-number-resolving detection is implemented \cite{npj2022, You2020Gerritsfication}. Finally, these detection events are recorded by a time tagger (PicoQuant MultiHarp 150) and analyzed. This experimental setup allows us to accurately measure the joint photon-number distribution at both outputs of the fiber beam-splitter. This enables us to perform photon subtraction and post-selection for image reconstruction.

\subsection{Data Analysis and Image Reconstruction}
In the compressive sensing process, we sequentially display some percentage of 4096 unique random matrices on the second DMD, with the measurement time for each matrix fixed at one second. For example, the reconstructions for Fig. \ref{fig:figure2} and \ref{fig:figure3} utilized 1025 projective measurements, which corresponds to 25\% of the 4096 measurements that our CS algorithm can use. The images reported on Fig. \ref{fig:figure4} were obtained with 12\% of the measurements, which means 512 binary matrices. To reconstruct the image, we apply the TVAL3 algorithm \cite{li2010efficient}. Due to the long coherence time of our pseudo-thermal light source, we bin each one-second measurement into 1 $\mu$s intervals \cite{You2020Gerritsfication, HashemiRafsanjani2017}. Then we count the number of detections in each bin, and this defines a photon-number resolving event. For each event, we then denote the photon-counts in the two arms as $n_1$ and $n_2$. To achieve $N$-photon subtraction, we isolate the events where $n_2 = N$. In other words, we filter $n_1$ by only considering the events where $n_2 = N$ in the second arm. We then perform the reconstruction process with this conditional dataset to obtain an enhanced image quality. Conversely, for post-selection on $n$ and $m$ photons, we compute the probability that $n_1=n$ and $n_2=m$. Then, we perform image reconstruction using this probability to obtain an improvement in image quality.

\section{ACKNOWLEDGEMENTS}
M.H., R.B.D., C.Y. and O.M.S.L. acknowledge funding from the US Department of Energy, Office of Basic Energy Sciences, Division of Materials Sciences and Engineering under Award DE-SC0021069. F.M. acknowledge funding from the National Science Foundation through Grant No. ECCS-2225986. R.J.L.-M. thankfully acknowledges financial support by DGAPA-UNAM under the project UNAM-PAPIIT IN101623. R. J. thanks support from the National Natural Science Foundations of China (Grant Numbers 92365106 and 12074299).

\section{COMPETING INTERESTS}
The authors declare no competing interests.

\section{DATA AVAILABILITY}
The data sets generated and/or analyzed during this study are available from the corresponding author or last author on reasonable request.

\bibliography{main}

\clearpage

\onecolumngrid

\section{Probability of observing multiphoton events}

In this section, we provide additional experimental results. In Table S1, we report the values associated with the probability of measuring specific multiphoton events from a thermal light beam. Specifically, Table S1 presents the probability of observing a particular number of photons under the post-selection scheme of Fig. 2. Similarly, Table S2 presents the joint probabilities of measuring a particular number of photons in two separate arms under the post-selection scheme of Fig. 3. Lastly, Table S3 presents the probability of observing a particular number of photons in the second arm under the photon-subtraction scheme of Fig. 4.

\begin{table}[!htbp]
\centering
\caption{The measured probability of post selection.}
\label{tab:table1}
\begin{tabular}{|p{1.5cm}<{\centering}|p{1.5cm}<{\centering}|p{1.5cm}<{\centering}|p{1.5cm}<{\centering}|p{1.5cm}<{\centering}|p{1.5cm}<{\centering}|p{1.5cm}<{\centering}|p{1.5cm}<{\centering}|p{1.5cm}<{\centering}|}
\hline
$\bar{n}$&$|0\rangle\langle0|$& $ |1\rangle\langle1|$ &$ |2\rangle\langle2| $ &$ |3\rangle\langle3|$& $|4\rangle\langle4|$& $ |5\rangle\langle5|$& $ |6\rangle\langle6|$& $ |7\rangle\langle7|$\\
\hline
$0.8 $&$55.17\%$ &$26.11 \%$ &$10.76\%$ &$4.51\%$& $ 1.95\%$& $ 0.87\%$ & $ 0.40\%$&$ 0.18\%$\\
\hline
\end{tabular}
\end{table}

\begin{table}[!htbp]
\centering
\caption{The measured probability of correlation between the two arm in the source.}
\label{tab:table1}
\begin{tabular}{|p{1.75cm}<{\centering}|p{1.5cm}<{\centering}<{\centering}|p{1.5cm}<{\centering}|p{1.5cm}<{\centering}|p{1.5cm}<{\centering}|p{1.5cm}<{\centering}|p{1.5cm}<{\centering}|p{1.5cm}<{\centering}|p{1.5cm}<{\centering}|p{1.5cm}<{\centering}|}
\hline
\diagbox{Arm1}{Arm2} & $|0\rangle\langle0|$& $ |1\rangle\langle1|$ &$ |2\rangle\langle2| $ &$ |3\rangle\langle3|$& $|4\rangle\langle4|$& $ |5\rangle\langle5|$& $ |6\rangle\langle6|$& $ |7\rangle\langle7|$\\
\hline
$|0\rangle\langle0|$&$15.99\%$ &$9.40 \%$ &$4.26\%$ &$1.78\%$& $ 0.73\%$& $ 0.28\%$ & $ 0.11\%$&$ 0.04\%$\\
\hline
$ |1\rangle\langle1|$ &$7.53\%$ &$7.05 \%$ &$4.64\%$ &$2.65\%$& $ 1.41\%$& $ 0.68\%$ & $ 0.33\%$&$ 0.14\%$\\
\hline
$ |2\rangle\langle2| $ &$2.70\%$ &$3.68 \%$ &$3.25\%$ &$2.39\%$& $ 1.54\%$& $ 0.92\%$ & $ 0.50\%$&$ 0.28\%$\\
\hline
$ |3\rangle\langle3|$&$0.89\%$ &$1.64 \%$ &$1.86\%$ &$1.70\%$& $ 1.33\%$& $ 0.94\%$ & $ 0.61\%$&$ 0.37\%$\\
\hline
$|4\rangle\langle4|$&$0.27\%$ &$0.65 \%$ &$0.93\%$ &$1.02\%$& $ 0.93\%$& 
$ 0.77\%$ & $ 0.57\%$&$ 0.40\%$\\
\hline
$ |5\rangle\langle5|$&$0.08\%$ &$0.24 \%$ &$0.42\%$ &$0.54\%$& $ 0.58\%$& 
$ 0.54\%$ & $ 0.46\%$&$ 0.36\%$\\
\hline
$ |6\rangle\langle6|$&$0.02\%$ &$0.08 \%$ &$0.18\%$ &$0.26\%$& $ 0.32\%$& 
$ 0.35\%$ & $ 0.34\%$&$ 0.29\%$\\
\hline
$ |7\rangle\langle7|$&$0.007\%$ &$0.02 \%$ &$0.06\%$ &$0.12\%$& $ 0.16\%$& 
$ 0.19\%$ & $ 0.21\%$&$ 0.20\%$\\
\hline
\end{tabular}
\end{table}

\begin{table}[!htbp]
\centering
\caption{The measured probability of photon subtraction.}
\label{tab:table2}
\begin{tabular}
{|p{1.5cm}<{\centering}|p{1.5cm}<{\centering}|p{1.5cm}<{\centering}|p{1.5cm}<{\centering}|p{1.5cm}<{\centering}|}
\hline
$\bar{n}$&$N=0$&$N=1$& $ N=2$ &$ N=3 $\\
\hline
$0.08 $&$97\%$ &$2.1 \%$ &$0.3\%$& $0.05\%$\\
\hline
\end{tabular}
\end{table}

\section{Detailed derivation of equations}

Here we provide a detailed derivation of the equations presented in the main body of our paper. The initial quantum state of our signal is a weak, single-mode thermal state. Written explicitly in the Fock basis, our initial thermal state of light is represented by 
\begin{equation}
    \hat{\rho}_{0} = \sum_{n=0}^\infty \frac{\bar{n}^n}{(1+\bar{n})^{n+1}}\left|n\rangle\langle n\right|,
\end{equation}
where $\bar{n}$ is the mean number of photons of the state. In our experiment, we uniformly illuminate an object with this state, producing a signal with a new state that has a different mode structure. The mode information is contained within the annihilation operator $\hat{a}$ which obeys $\hat{a}|n\rangle = \sqrt{n}|n-1\rangle$, defined in terms of the operator-valued distribution $\hat{a}(\Vec{\boldsymbol{x}})$ by
\begin{equation}
    \hat{a} = \int d^2 x f(\Vec{\boldsymbol{x}})\hat{a}(\Vec{\boldsymbol{x}}),
\end{equation}
where $\left[\hat{a}(\Vec{\boldsymbol{x}}),\hat{a}^\dagger(\Vec{\boldsymbol{x}}')\right] = (2\pi)^2\delta(\Vec{\boldsymbol{x}}-\Vec{\boldsymbol{x}}')$ is the canonical commutation relation and $f(\Vec{\boldsymbol{x}})$ is the transverse profile of the beam. This expression assumes that the light is strongly peaked around a particular frequency, and in this case a transverse positional description can be used.

In our experiment, we uniformly illuminate an object using the thermal state, which forms an image that we would like to measure. We do this by discretizing the transverse spatial profile of the mode into $X$ squares which we will call pixels. This is equivalent to the transformation taking $\hat{a}$ to $\sum_{i=1}^X \lambda_i \hat{A}_i$ where $\hat{A}_i$ is the annihilation operator for the mode at the $i^{\text{th}}$ pixel and $\lambda_i$ is its weight. Since the object was illuminated uniformly, $\lambda_i$ will be either $0$, representing a pixel with no light, or some constant value, representing a pixel with light, such that $\sum_{i=1}^X|\lambda_i|^2 = 1$. It is important to note, however, that this theory will also apply for non-uniform illuminations. This allows us to define the ideal image vector $\Vec{\boldsymbol{s}}_0\in \mathbb{R}^X$ where each component $s_{0,i}$ is equal to $|\lambda_i|^2\bar{n}$.

We now collect random combinations of these pixels onto a single-pixel camera that employs photon-number-resolving detection. We will see later how this allows for image reconstructions which use fewer measurements than traditional methods require. We will perform $M$ such measurements, and each random selection of pixels will be represented by the covector $\Vec{\boldsymbol{Q}}_t\in\mathbb{R}^{X*}$ which consists of zeros and ones. It follows that, after the signal has been filtered by this covector, the resulting mode operator of the signal will be given by $\hat{a}_t=\sum_{i=0}^X Q_{t,i}\lambda_i' \hat{A}_i$ where $\lambda_i' = \lambda_i/\sqrt{\sum_{i=0}^\infty Q_{t,i}|\lambda_i|^2}$ is the re-normalized weight of each pixel. The quantum state of the signal after this filtering process is therefore thermal, with a mean-photon-number given by $\bar{n}_t = \Vec{\boldsymbol{Q}}_t\cdot \Vec{\boldsymbol{s}}_0$, and can be written as
\begin{equation}
    \hat{\rho}_{\boldsymbol{Q},t} = \sum_{n=0}^\infty \frac{\bar{n_t}^n}{(1+\bar{n}_t)^{n+1}}\left|n\rangle\langle n\right|.
\end{equation}
Here we are using the label $\boldsymbol{Q}$, which represents the matrix of pixel filtrations and is defined by $\boldsymbol{Q} = \bigoplus_{t=1}^M \Vec{\boldsymbol{Q}}_t$. We can simultaneously write all such density matrices as
\begin{equation}
    \boldsymbol{\hat{\rho}}_{\boldsymbol{Q}} = \bigoplus_{t=1}^M \hat{\rho}_{\boldsymbol{Q},t},
\end{equation}
which can be thought of as a vector of density matrices. 

We now use the Gluaber-Sudarshan $P$ function representation of the quantum state, written in terms of coherent states $|\alpha\rangle$, and given by
\begin{equation}
    \boldsymbol{\hat{\rho}}_{\boldsymbol{Q}} = \bigoplus_{t=1}^M \int d^2\alpha \frac{1}{\pi \bar{n}_t}e^{-\frac{|\alpha|^2}{\bar{n}_t}}\left|\alpha\rangle\langle\alpha\right|.
\end{equation}
Before measuring the state with our photon-number-resolving detector, we will send it through a fiber-coupler in order to produce a second mode that can be used for the photon-subtraction technique which we will discuss later. After this transformation, represented by taking annihilation operator $\hat{a}_t$ to annihilation operators $\hat{a}_t\cos(\theta) + i \hat{b}_t\sin(\theta)$ where $\theta$ is the beam-splitter angle, the state is given by
\begin{equation}
    \boldsymbol{\hat{\rho}}_{\boldsymbol{Q}} = \bigoplus_{t=1}^M \int d^2\alpha \frac{1}{\pi \bar{n}_t}e^{-\frac{|\alpha|^2}{\bar{n}_t}}\left|\alpha\cos(\theta),i\alpha\sin(\theta)\rangle\langle\alpha\cos(\theta),i\alpha\sin(\theta)\right|_{a,b}.
\end{equation}
We will use the labels $a,b$ to represent the two output modes of the fiber-coupler.

From here, we make use of two photon-number-resolving detectors, one in each arm, to perform measurements. The primary difficulty of this measurement scheme is that the signal's strength is comparable to the noise of our two detectors, and the measurement techniques which we will employ are meant to alleviate the effects of that noise. The impacts of noise and detector efficiencies can be modeled with the photocounting technique, by which for a given state $\hat{\rho} = \sum_{n=0}^\infty p(n,m)\left|n\rangle \langle m\right|$, its diagonal matrix elements $p_{\text{noise}}(n,n)$ with dark counts $\nu$ and detector efficiency $\eta$ accounted for can be computed by
\begin{equation}
    p_{\text{loss}}(n,n) = \left\langle:\frac{(\eta\hat{n}+\nu)^n}{n!}e^{-(\eta\hat{n} + \nu)}:\right\rangle,
\end{equation}
where $:\cdot:$ is the normal-ordering prescription. In our case, these diagonal elements can be computed for dark counts $\nu_{a/b}$ and detector efficiencies $\eta_{a/b}$ as
\begin{equation}
    \begin{aligned}
               \Vec{\boldsymbol{p}}_{\boldsymbol{Q}}(n, m)&=\bigoplus_{t=1}^M\left\langle: \frac{\left(\eta_a \hat{n}_a+\nu_a\right)^n}{n !} e^{-\left(\eta_a \hat{n}_a+\nu_a\right)} \otimes \frac{\left(\eta_b \hat{n}_b+\nu_b\right)^m}{m !} e^{-\left(\eta_b \hat{n}_b+\nu_b\right)}:\right\rangle\\
               &=\bigoplus_{t=1}^M \int d^2\alpha \frac{1}{\pi \bar{n}_t}e^{-\frac{|\alpha|^2}{\bar{n}_t}} \frac{\left(\eta_a |\alpha|^2\cos^2(\theta)+\nu_a\right)^n}{n !} e^{-\left(\eta_a |\alpha|^2\cos^2(\theta)+\nu_a\right)}\frac{\left(\eta_b |\alpha|^2\sin^2(\theta)+\nu_b\right)^m}{m !} e^{-\left(\eta_b |\alpha|^2\sin^2(\theta)+\nu_b\right)} \\
               &= \bigoplus_{t=1}^M\frac{e^{-\nu_a-\nu_b}}{n! m!}\sum_{i=0}^n\sum_{j=0}^m \binom{n}{i}\binom{m}{j}\eta_a^i\eta_b^j\nu_a^{n-i}\nu_b^{n-j}\cos^{2i}(\theta)\sin^{2j}(\theta)\int d^2\alpha \frac{|\alpha|^{2i+2j}}{\pi \bar{n}_t}e^{-\frac{|\alpha|^2}{\bar{n}_t} - \eta_a|\alpha|^2\cos^2(\theta) - \eta_b|\alpha|^2\sin^2(\theta)} \\
               &=\bigoplus_{t=1}^M\frac{e^{-\nu_a-\nu_b}}{\bar{n}_t n! m!}\sum_{i=0}^n\sum_{j=0}^m \binom{n}{i}\binom{m}{j}(i+j)!\frac{\eta_a^i\eta_b^j\nu_a^{n-i}\nu_b^{m-j}}{\left(\frac{1}{\bar{n}_t}+\eta_a\cos^2(\theta)+\eta_b\sin^2(\theta)\right)^{1+i+j}}\cos^{2i}(\theta)\sin^{2j}(\theta).
   \end{aligned}
\end{equation}
Unfortunately, the finite sum in the last line of this expression does not have a nice analytical form. However, since it is a finite sum, these diagonal matrix elements can be easily calculated numerically. 
When the signal $\bar{n}_t$ is absent, we will detect only the noise. In this case, the joint probability of noise event is given by $p_n(k,l) = p_{n,a}(k)p_{n,b}(l)$, where $p_{n,i}(k) = e^{-\nu_i}\frac{\nu_i^k}{k!}$. We note that our ability to reliably reconstruct the signal's mode profile from our measurements hinges on each of the $M$ measurements in arm $a$ being clearly distinguishable from its background noise. In other words, the signal-to-noise ratio for each measurement should be as high as possible. Let us now discuss two methods for accomplishing this.

The first method is that of post-selection (Fock-projection) in arm $a$. This method does not utilize arm $b$, so that arm will always be traced out here. The signal-to-noise ratio in the case where we post-select on $N$ photons in arm $a$ can be represented by a vector, and is written as
\begin{equation}
    \overrightarrow{\textbf{SNR}}_{\text{post}}(N) = \frac{\sum_{m=0}^\infty \Vec{\boldsymbol{p}}_{\boldsymbol{Q}}(N,m)}{p_{n,a}(N)}=\frac{\Vec{\boldsymbol{p}}_{\boldsymbol{Q}}(N)}{p_{n,a}(N)}.
\end{equation}
Numerical evaluations of this quantity show that each component of the signal-to-noise ratio vector is increasing in an approximately exponential fashion with respect to $N$. This shows that we can greatly reduce the impact of noise on our data by post-selecting on high photon numbers.


The other method showcased in this paper is that of photon-subtraction, by which we first make a post-selective measurement in arm $b$ on $N$ photons and then measure the photon events in arm $a$. The conditional intensity in arm $a$ can then be written as $\langle\hat{\boldsymbol{n}}_a\rangle_N = \bigoplus_{t=0}^M\left(\sum_{k=0}^\infty k p_{\boldsymbol{Q},t}(k,N)\right)/\left(\sum_{k=0}^\infty p_{\boldsymbol{Q},t}(k,N)\right)$, where the factor in the denominator is due to the renormalization of the state after the measurement in arm $b$. Similarly, the noise measurement can be written as $\langle\hat{n}_a\rangle_{N,0} = \bigoplus_{t=0}^M\left(\sum_{k=0}^\infty k p_n(k,N)\right)/\left(\sum_{k=0}^\infty p_n(k,N)\right)$. By taking this approach, the resulting signal-to-noise ratio seen in arm $a$ can be represented by
\begin{equation}
    \overrightarrow{\textbf{SNR}}_{\text{sub}}(N) = \frac{\langle\hat{\boldsymbol{n}}_a\rangle_N}{\langle\hat{n}_a\rangle_{N,0}}.
\end{equation}
In contrast to the post-selection case, each component in this vector increases in an approximately linear fashion with respect to $N$. While this may be less desirable when compared to the exponential trend of post-selection in arm $a$, it is useful when precise post-selective measurements in arm $a$ cannot be made. For instance, if $\bar{n}_t$ is very large, then we can choose $\theta$ to be very small so that photon-number-resolution can be made accurate in arm $b$. This would allow us to increase the signal-to-noise ratio in arm $a$ through photon subtraction while making the more-precise measurement of intensity in that arm.

Finally, our measurements in arm $a$ will be used to form a reconstruction of the signal vector, $\Vec{\boldsymbol{s}}_0$. This is accomplished using the compressive sensing (CS) technique, by which the reconstructed image, represented by $\Vec{\boldsymbol{s}}\in \mathbb{R}^X$, is found by minimizing the following quantity with respect to the dummy-vector $\Vec{\boldsymbol{s}}'\in \mathbb{R}^X$:
\begin{equation}
     \sum_{i=0}^X \lVert\nabla s_i'\rVert_{l_1} + \frac{\mu}{2}\lVert \boldsymbol{Q}\Vec{\boldsymbol{s}}' - \langle\hat{\boldsymbol{n}}\rangle\rVert_{l_2}.
\end{equation}
Here, $\langle\hat{\boldsymbol{n}}\rangle$ could be replaced with either of the previously-described quantities,  $\Vec{\boldsymbol{p}}_{\boldsymbol{Q}}(N)$ or $\langle\hat{\boldsymbol{n}}_a\rangle_N$. Moreover, the $1$- and $2$-norm are denoted by $\lVert\cdot\rVert_{l_1}$ and $\lVert\cdot\rVert_{l_2}$, respectively. The discrete gradient operator is described by $\nabla$, and the penalty factor by $\mu$. The value of $\Vec{\boldsymbol{s}}'$ which minimizes this quantity is then the value which we ascribe to $\Vec{\boldsymbol{s}}$. Accurate reconstruction of this image vector, such that $\Vec{\boldsymbol{s}}$ agrees with $\Vec{\boldsymbol{s}}_0$, is sensitive to background noise, and so by reducing the impact of that noise as much as possible via either of the two methods described above, we can attain a more reliable image of the signal.

\end{document}